\documentclass[journal]{vgtc}                     

\onlineid{0}

\vgtccategory{Research}

\vgtcpapertype{please specify}

\title{\texorpdfstring{Designing a Visualization Atlas: Lessons \& Reflections\\ from The UK Co-Benefits Atlas for Climate Mitigation}{Designing a Visualization Atlas: Lessons \& Reflections from The UK Co-Benefits Atlas for Climate Mitigation}}

\author{%
  \authororcid{Jinrui Wang}{0009-0007-9313-5180},
  \authororcid{Alexis Pister}{0000-0002-2817-020X},
  \authororcid{Sian Phillips}{0009-0001-4006-5525},
  \authororcid{Sarah Bissett}{},
  \authororcid{Ruaidhri Higgins-Lavery}{0009-0001-0117-0541}, \texorpdfstring{\\}{}
  \authororcid{Clare Wharmby}{},
  \authororcid{Andrew Sudmant}{0000-0001-8650-8419},
  \authororcid{Uta Hinrichs}{0000-0001-7494-0941},
  \authororcid{Benjamin Bach}{0000-0002-9201-7744}
}

\authorfooter{
  \item
  	Jinrui Wang, Sarah Bissett, Ruaidhri Higgins-Lavery, Clare Wharmby, Andrew Sudmant, and Uta Hinrichs are with The University of Edinburgh.
    E-mail: \href{mailto:jinrui.wang@ed.ac.uk}{jinrui.wang},
    \href{mailto:Sarah.Bissett@ed.ac.uk}{sarah.bissett},
    \href{mailto:r.higgins-lavery@ed.ac.uk}{r.higgins-lavery},
    \href{mailto:Clare.Wharmby@ed.ac.uk}{clare.wharmby},
    \href{mailto:andrew.sudmant@ed.ac.uk}{andrew.sudmant},
    \href{mailto:uhinrich@ed.ac.uk}{uhinrichs}@ed.ac.uk.
  \item 
    Alexis Pister and Sian Phillips are with City St George's, University of London.
    Email: \href{mailto:alexis.pister@city.ac.uk}{alexis.pister},
    \href{mailto:sian.phillips@city.ac.uk}{sian.phillips}@city.ac.uk.
  \item
  	Benjamin Bach is with Inria Bordeaux.
  	E-mail: \href{mailto:benjamin.bach@inria.fr}{benjamin.bach@inria.fr}
}

\abstract{%
  This paper reports on the process of designing the UK Co-Benefits Atlas, which communicates and publicizes data for climate mitigation. Visualization atlases—an emerging type of platform to make data about complex topics comprehensive through interactive visualizations and explanatory content—pose challenges beyond traditional visualization projects. Atlases must address diverse and often uncertain audiences and use cases, support both explanatory and guided exploration, and accommodate complex, evolving data. 
Over 10 months, our team of visualization and domain experts conducted 8 design workshops, iterative prototyping, 15 stakeholder onboarding sessions, and continuous reflection. These intertwined processes informed the development of the Atlas, comprising over 400 pages of visualizations and explanations. They also enabled a deeper understanding of how stakeholders may critically engage with the atlas in practice, in terms of interests, potential frictions when navigating huge amounts of data, and envisioned usage scenarios.
Reflecting on our design process, we identify five driving forces in atlas design—data, people, stories, context, and the atlas itself—whose shifting dynamics influence different stages of visualization atlas design in different ways. 
Grounded in our case study, we discuss using these forces as a conceptual starting point for structuring and reflecting on future atlas design processes.

}


\teaser{
  \centering
  \includegraphics[width=\linewidth, alt={The figure shows an overview of the UK Co-Benefits Atlas interface. It displays five types of pages: the landing page, 11 co-benefit report pages, 386 local authority pages, 17 socio-economic factor pages, and an interactive map. Each page is a customized template of charts, maps, and explanations. Below the screenshots, text summarizes the people-centered design approach through workshops, design decisions, and stakeholder engagement, and lists the people involved, including visualization experts, domain experts, and more than 200 external stakeholders.}]{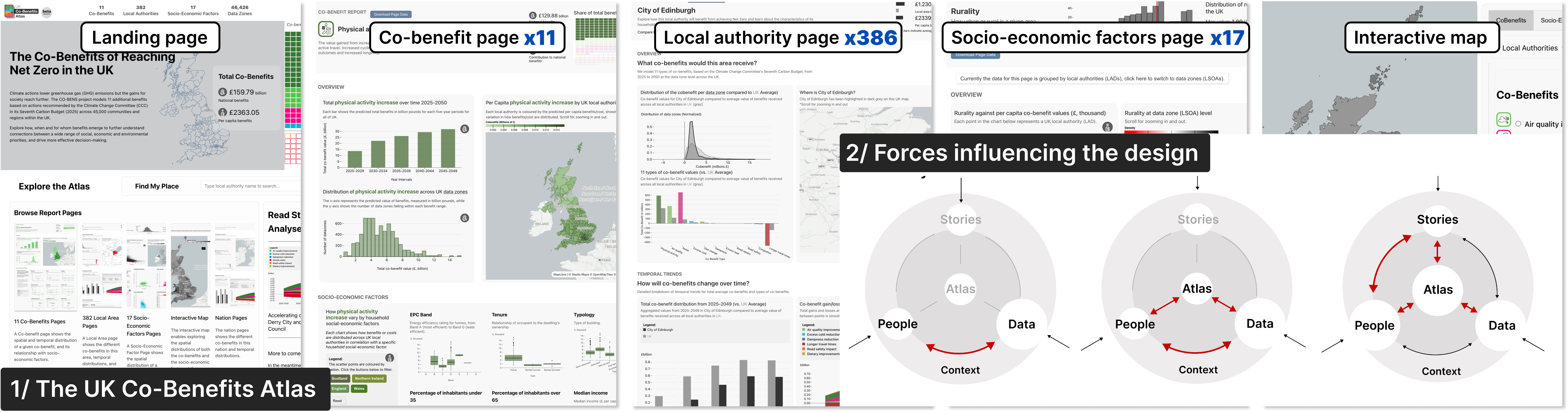}
  \caption{%
  	Overview of the UK Co-Benefits Atlas: (from left) Landing Page, Co-Benefit Report Pages, Local Authority Pages, Socio-Economic Factors, and Interactive Map. The design process led us to identifying five major \textit{forces} that influence the design. 
  }
  \label{fig:teaser}
}

\graphicspath{{figs/}{figures/}{pictures/}{images/}{./}} 

\usepackage{tabu}                      
\usepackage{booktabs}                  
\usepackage{lipsum}                    
\usepackage{mwe}                       

\usepackage{mathptmx}                  
\usepackage{xspace}
\usepackage{url}
\usepackage{wrapfig}
\usepackage{multirow}
\usepackage{tabularray}
\usepackage{longtable}
\usepackage{soul}
\usepackage{xtab}

\usepackage{tikz}
\usepackage{xcolor}
\usepackage{marginnote}
\begin{document}



\newcommand{\jumpto}[1]{%
  \href{https://docs-ukcobenefitsatlas.github.io/\##1}{[#1]}
}

\newcommand{\itquote}[1]{\textit{``#1''}}

\maketitle

\section{Introduction}

Visualization atlases are an emerging genre of open data platforms designed to make complex datasets about societal challenges available, transparent, and  comprehensive~\cite{wang_visualization_2025}.
Prominent examples include the Atlas of Economic Complexity~\cite{GrowthLab2013Atlas}, the Atlas of UN Sustainable Development Goals~\cite{atlasSustainability}, and the IPCC's WGI Interactive Atlas of possible climate futures~\cite{atlasIPCC}. While all of these examples provide comprehensive multidimensional data from public sources and institutions, they also go beyond traditional open-data platforms that merely grant access to data files: these atlases leverage interactive visualizations and rich contextual explanations to curate cohesive collections of visualizations, dashboards, reports, and stories. They also typically offer onboarding, navigation, and methodological explanations to support their readers.

Providing multiple means of engaging with data is complex, and visualizations for exploration and explanation introduce challenges not captured by existing visualization design methodologies (e.g., \cite{sedlmair2012design, oppermann2020data,munznerNestedModelVisualization2009}).
Yet, the limited written accounts on atlas design~\cite{goodwin2024designing,10.1145/3544549.3573866, oppermann_bike_2018} provide some early, yet incomplete evidence of the challenges inherent to larger atlas projects. These challenges to atlas design typically include large and multidimensional datasets and require extensive resources for design, development, and communication. Visualization atlases are also designed for diverse audiences and different use cases: they generally aim to inform, educate, and support both in-depth analysis and decision making. 
Atlases need to address an audience spanning domain experts, researchers, businesses, educators, and the public.
Each of these user groups has different objectives and knowledge about the topic in question, the visualizations provided, and the ways in which data have been collected and analyzed, requiring careful explanations and onboarding. Existing visualization design guidelines and frameworks are generally directed towards specific forms of visualization (e.g., dashboards \cite{bach2022dashboard}, data-stories \cite{lee2015more}, news articles \cite{shan2024design}) with rather specific tasks and user groups.
Ultimately, atlases need to dynamically adapt to changes in data and topics they present~\cite{hinrichs_defense_2019, vines_configuring_2013, walny_data_2020}.

To better understand these design challenges and provide guidance on future visualization atlas projects, we present a case study on designing an atlas platform in the context of climate mitigation. The atlas is built on a dataset of modeled projections of future monetary gains (benefits) arising from climate interventions to reduce CO$_2$ emissions. For example, such interventions may lead to reduced healthcare costs due to improved air quality or lower energy bills through better home insulation, etc.~\cite{sudmant2024climate}. Among others, these co-benefits data are intended to support policy makers in planning and advocating for climate mitigation projects and provide data to researchers doing refined climate mitigation and economic analyses. 
It is the first dataset of its kind, covering 46,426 local communities, 11 co-benefits, 25 years of projected data, and 17 socio-economic factors (including education, income, household size, etc). Over a period of 10 months, our team of visualization designers and domain experts in policy and climate mitigation led the co-design and implementation of the resulting \textit{The UK Co-Benefits Atlas} (\url{https://ukcobenefitsatlas.net}).

Based on experiences, findings and reflections from eight~co-design workshops, 15 sessions with external stakeholders, and prototype iterations, we provide the first comprehensive account of visualization atlas design. This includes: (1) a documented design process featuring interdisciplinary collaboration and design workshops (\cref{sec:methodology}); (2) a set of bespoke design solutions comprising multiple page templates and structural components (\cref{sec:atlas}); and (3) insights from stakeholders regarding their first impressions, frictions introduced by certain Atlas features, and envisioned usage scenarios (\cref{sec:stakeholders}).
We reflect on our design process through the metaphor of five distinct forces we identified, which influence each other and the design process: \textsc{data}, \textsc{people}, \textsc{stories}, \textsc{context}, and the \textsc{atlas} itself (\cref{sec:discussion}). We discuss how these five forces create a `spatial', rather than temporally linear approach to atlas design, implying no single starting or end point, and instead suggesting a dynamic, non-linear, and potentially never-ending design process. This forces metaphor also allows for linking existing design and evaluation methods into an atlas project, without prescribing one singular methodology. We conclude by outlining future avenues (\cref{sec:conclusion}) to explore in the context of visualization atlas design.

\section{Background}
\label{sec:background}

Our work builds on recent research on visualization atlases as a relatively new form of visualization. It is further grounded in visualization design methodology on the one hand and the specifics of our case study---climate co-benefits modeling---on the other, as outlined below.

\subsection{Visualization Atlases}

Wang et al.~\cite{wang_visualization_2025} define the concept of a visualization atlas as \textit{``a compendium of (web) pages aimed at explaining and supporting the exploration of data about a dedicated topic through data, visualizations and narration''.} This definition highlights three main characteristics of visualization atlases: Firstly, their focus on a \textbf{topic of public interest} but of considerable complexity (e.g., economic inequality and development~\cite{GrowthLab2013Atlas}, sustainability~\cite{atlasSustainability,atlasIPCC} or public health~\cite{goodwin2024designing}), and their aim of providing data-driven views that support a broad understanding of this topic. Secondly, visualization atlases leverage \textbf{interactive visualization and narrative capabilities} to let users explore the data while providing sufficient explanatory power to guide and inform. They can incorporate all forms of visualization; interactive charts, dashboards~\cite{bach2022dashboard}, data videos~\cite{10.1145/2702123.2702431}, data-articles~\cite{shan2024design}, and more. 
Unlike traditional exploratory or explanatory visualizations, atlases do not assume explicit audiences, goals, or tasks. Instead, they are designed to support users who are already familiar with the topic and seeking access to data, those wanting to learn about it, and those interested in scrutinizing or extending it.
Third and finally, atlases are highly \textbf{curated sets of bespoke pages} that consist of visualizations, explanations, and guidance, navigable through a clear structure. Curation is essential to increasing communication and accessibility, while also framing the dataset, its purpose, and potential use.
To that end, Wang et al. identified \textit{design patterns} for specific design decisions such as entry page design, atlas structure, or visualization/interaction designs, and atlas \textit{genres} as distinct approaches to presentation and engagement—for example, Storybanks provide sets of written stories and in-depth analyses; Observatories track and monitor multiple datasets; Exploratoria emphasize interactive exploration. While these patterns, genres, and examples demonstrate the sophistication and diversity of atlas designs, they give little actionable indication of the actual design process and how to navigate specific problems of data complexity and user engagement in practice. Our goal is to better understand the dynamics of atlas design processes and decision-making as part of this in order to scaffold the design of future atlases.

\subsection{Approaches to Visualization Design}
\label{sec:methodologies}

A rich body of work exists on visualization methodology, focused in particular on scaffolding visualization design processes.   
Some of the most prominent examples include the design activity framework of understand, ideate, make, and deploy~\cite{mckenna_design_2014}; the classic nine-step process of visualization design studies~\cite{sedlmair2012design}; and the extension towards data-first design studies~\cite{oppermann2020data}. Across these models, task abstraction plays a central role, supported by frameworks such as the nested model~\cite{munznerNestedModelVisualization2009} and the multi-level typology~\cite{brehmer2013multi}. While these models are widely applicable, most of these frameworks are built on specific assumptions such as a well-defined problem and data, a set of constraint stakeholders that can be analyzed, as well as a clear linear or iterative \textit{progression} towards a visualization outcome. Moreover, these models assume that this process is \textit{deterministic} with a measure for when to stop.

However, a growing body of visualization research has begun to adopt concepts from the wider design research community that promote co-creation, reflective, and non-linear methodology.
These considerations include design-by-immersion that promote transdisciplinary engagement~\cite{hall_design_2020}, applying action design research in visualization studies~\cite{mccurdy_action_2016}, and applying an interpretivist perspective toward design studies~\cite{meyer_criteria_2019, rogers2020insights}. More recent perspectives suggest that both problem and visualization co-evolve in designers' practice~\cite{parsons2025beyond}, echoing broader design research on creativity~\cite{dorst2001creativity} and problem--solution co-evolution~\cite{sanders2008co}. Similarly, many (co-)design processes involve a ``fuzzy front end'' and are highly non-linear and perhaps non-deterministic \cite{sanders2008co,wei2026fuzzy}.
Further work addresses specific aspects within a design cycle, for example, structured visualization workshops to surface design opportunities~\cite{kerzner_framework_2019}, and accounts of evolving data and handover between visualization designers and engineers~\cite{walny_data_2020}. 
Building on these perspectives, we aim to explore a more flexible interpretation of existing design guidelines to better navigate the complexity of visualization atlas design.

\subsection{Case Study Context: Climate Co-Benefits Modeling}
We situate our inquiry into atlas design on a case study of climate co-benefits research in the UK.
Co-benefits refer to the wider socio-economic effects of reducing greenhouse gas emissions~\cite{karlsson2020climate}. For example, switching from fossil-fuel cars to electric vehicles results in reduced noise as well as improved air quality. 
Co-benefits can be expressed as monetary values representing money saved~\cite{dinh2024measuring}. They, therefore, can play a crucial role in closing the gap between the investments required for climate action and the perceived limited immediate payoffs. 
In total, our datasets contains \textbf{11~\textit{projected} co-benefits} such as air quality improvements, physical activity increase, and reduced dampness.
These projected co-benefits are based on national climate interventions proposed by the UK Climate Change Committee (CCC)'s Seventh Carbon Budget~\cite{ccc2025seventhcarbonbudget},  including improving building insulation, shifting to low-carbon modes of transport, and reducing meat consumption in favor of healthier diets, etc. 
They are modeled projections on a yearly basis from 2025 to 2050, which is the UK Net Zero target year, highlighting the timeliness of designing a visualization atlas for this context.  
Besides such \textit{temporal aspects}, the co-benefits data contains \textbf{17 source indicators} collected from census records and national surveys, encompassing \textit{continuous data} (e.g., median income), \textit{discrete data} (e.g., number of cars), and \textit{categorical data} (e.g., energy efficiency bands). All co-benefits values are modeled on a \textit{hyper-local} level, totaling \textbf{46,426 small geographical units} (Data Zone) across the entire UK.
Our case study dataset is developed with recognized UK government policy appraisal methods~\cite{HMTreasury2020GreenBook, frontier2022nzdm, CCC_2025_distribution_co_benefits} and informed by Sudmant et al.~\cite{sudmant2024climate}. Full details on methodology decisions, co-benefits selection, and data model optimization are available via the Atlas methodology page.

\begin{figure*}[!ht]
  \centering
  \includegraphics[width=\textwidth]{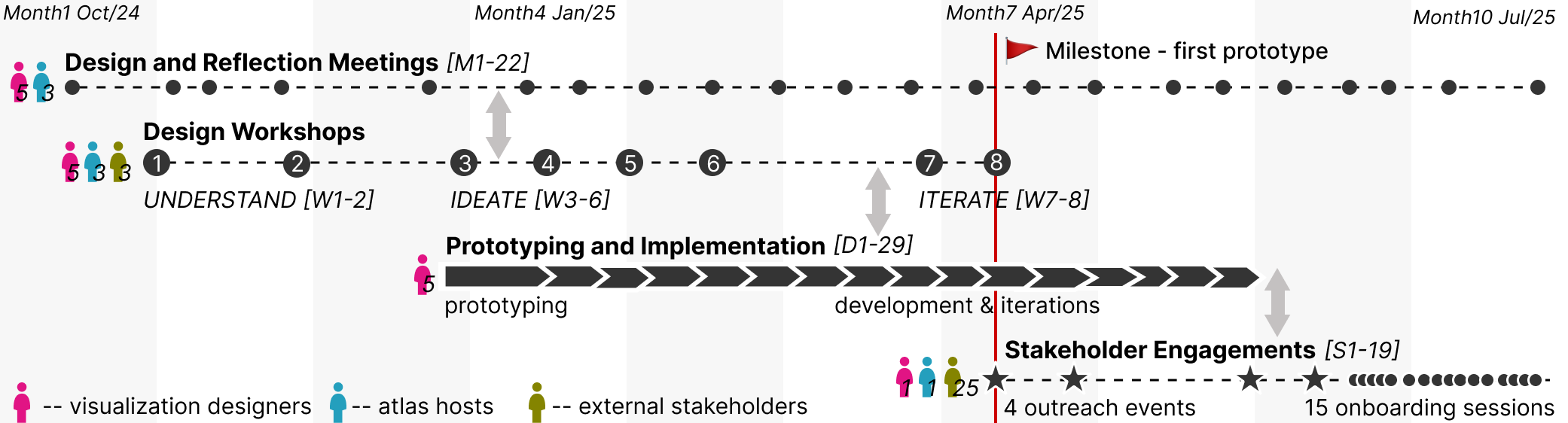}
  \caption{The UK Co-Benefits Atlas design process consisted of design and reflection meetings, workshops, prototyping \& development, and stakeholder engagements. The ``first prototype'' milestone marks the start of wider external stakeholder engagements.}
  \label{fig:project-overview}
\end{figure*}

\section{Co-Benefits Atlas: Research \& Design Methods}
\label{sec:methodology}
Our case study is motivated by the goal of creating a visualization atlas to make co-benefits data public and usable for diverse audiences. 
The novelty and multidimensionality of the data, combined with the \textit{wickedness} \cite{rittel1973dilemmas,levin2012overcoming} of the topic---climate change and mitigation---and its broad societal relevance, make co-benefits an ideal example to study visualization atlas design. Our goal was to visualize the entire data set, encompassing over 46,000 local communities, multiple data dimensions, and projected changes over time. In doing so, we aimed to learn from existing atlas projects, design patterns for exploration and explanation \cite{wang_visualization_2025}, and sought to better understand the challenges in designing open data platforms. In this section, we describe the people involved in the UK Co-benefits atlas design as well as the design activities and research methodologies for understanding the people, their potential goals and tasks, and eventually inform our design.

\subsection{People involved}
\label{sec:people-involved}

We describe the three groups of people involved in the design process by their relationship to the atlas:

\begin{itemize}[noitemsep,leftmargin=*, topsep=0pt, partopsep=0pt]
    \item \textbf{5 visualization designers} contributed expertise to visualization design, technical development, and research, three of whom have previously created a comprehensive survey and classification of atlases, including design patterns  \cite{wang_visualization_2025}. 
    We call them \textbf{designers} where two members primarily led the technical development, the remaining three focused on design concepts of pages and visualizations, planning and facilitating design workshops (see \cref{sec:workshops}), and coordinating stakeholder onboarding (see \cref{sec:stakeholders}).

    \item \textbf{3 atlas hosts} with a background in socio-economic analysis drove the co-benefits research~\cite{sudmant2024climate}, dataset modeling and analysis, and policy communication. 
    We call them \textbf{hosts} because they initiated the atlas topic, provided the dataset and findings, and facilitated a broader stakeholder engagement through public events.

    \item \textbf{28 external stakeholders} with diverse professional backgrounds spanning academia, public services, charities, and local communities. These stakeholders contributed valuable perspectives as potential atlas audiences. Three of these stakeholders participated in design workshops, with one attending multiple workshops. The remaining 25 stakeholders were consulted in later design stages to provide detailed feedback on the Atlas (\cref{sec:stakeholders}).
    
\end{itemize}

The authors of this paper comprise all the Atlas designers and hosts, along with one key stakeholder with long-term involvement in co-benefits policy work. All authors were consistently involved throughout the entire project and are therefore referred to as the \textit{core team}, capturing their collective insights and reflections.

\subsection{Research and Design Methods}

The Atlas design and research process was structured around four complementary strands of activities shown in \cref{fig:project-overview}: design and reflection meetings throughout the project, design workshops, prototyping and implementation, and stakeholder engagement through onboarding sessions. These strands offer complementary perspectives and collectively informed the Atlas design process.
The lead author documented individual activities and compiled them into an interactive timeline, hosted on the project website (\url{https://docs-ukcobenefitsatlas.github.io}).
This timeline records each activity’s agenda, key findings, and participants. Activities are referenced throughout the paper using hyperlinks, where each label (e.g., \jumpto{W1} for Workshop 1) links to the corresponding entry and associated materials in the timeline.
We outline the four strands of design and research activities below.

\begin{enumerate}[leftmargin=*, noitemsep, topsep=0pt, partopsep=0pt]
    \item \textbf{8 Design workshops} (\jumpto{W1}-\jumpto{W8}) aimed to collectively articulate atlas design goals, anticipate stakeholder needs, and explore design options across different components. The individual workshops spanned diverse topics in visualization \jumpto{M1}, inspired by Wang et al's observations on other atlases~\cite{wang_visualization_2025}. This list was then gradually refined over the course of the project, based on findings from previous workshops and emerging project priorities (e.g., understanding stakeholders; focus on data stories). 
    
    \item \textbf{Atlas prototype design \& development} (\jumpto{D1}-\jumpto{D29}) led by the visualization experts. Design started with sketches in~\jumpto{W2}, then prototyping began with two synchronous brainstorming sessions (\jumpto{D1}, \jumpto{D2}), followed by iterative development that resulted in the current Atlas prototype (\cref{sec:atlas}). The design process was reviewed and refined through design meetings, for example, deliberating different layouts of the co-benefit page (e.g.,~\jumpto{D19}) and considering different chart types for the co-benefits distribution (e.g.,~\jumpto{D15}). 
    
    \item \textbf{Stakeholder engagement} consisted of four public outreach events (\jumpto{S1}-\jumpto{S4}) and 15~in-depth onboarding sessions (\jumpto{S5}-\jumpto{S19}) with the initial working prototype (\cref{fig:project-overview}-milestone). These activities aimed to gather early feedback from stakeholders, including potential usage scenarios. Detailed methods and findings from the onboarding sessions are reported in \cref{sec:stakeholders}.
    
    \item \textbf{Design \& reflection meetings} (\jumpto{M1}-\jumpto{M22}) refer to discussions among the authors throughout the design process, from early stages to post-engagement. These meetings centered around workshop findings (e.g., discussing potential stakeholder roles~\jumpto{M2}), synchronizing design updates, tracking changes (e.g., major data updates~\jumpto{M12}), and planning subsequent activities and issues to discuss with the hosts (e.g., preparing data for sketching workshops~\jumpto{M4}). This reflective practice provided a consistent means for revisiting our goals and synthesizing findings.
\end{enumerate}

\begin{figure*}[ht]
  \centering
  \includegraphics[width=\textwidth]{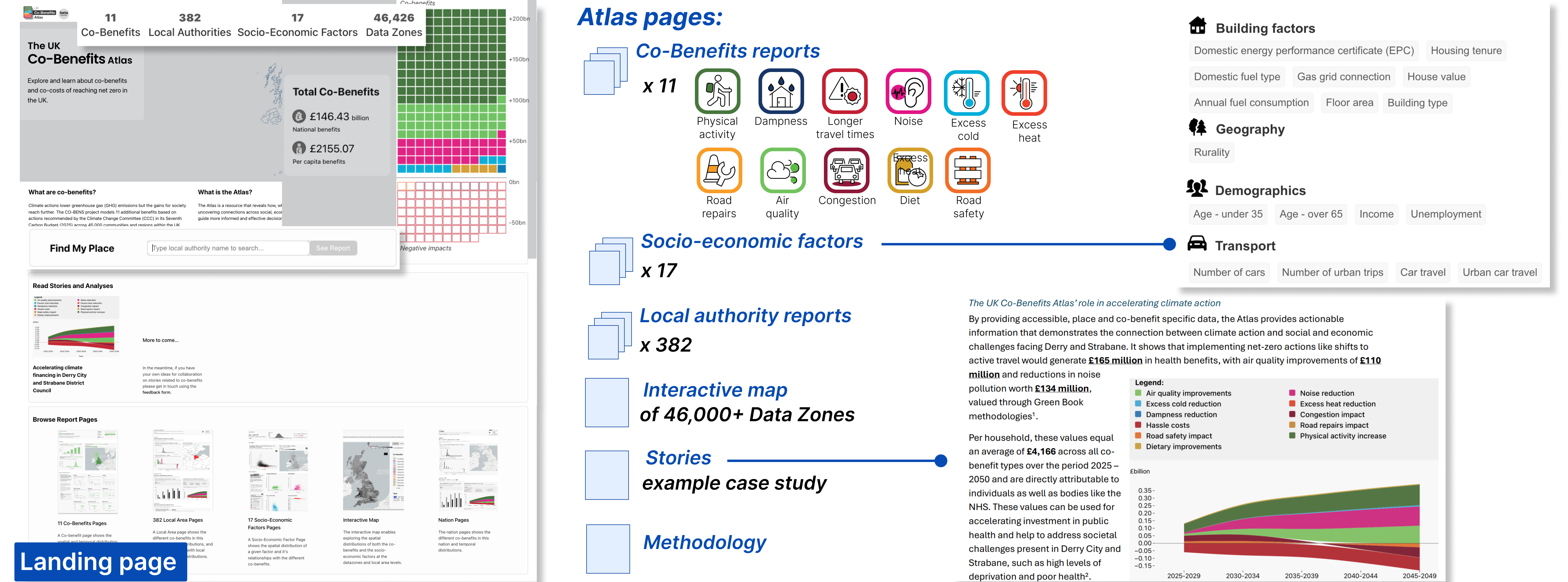}
  \caption{Overview of The UK Co-Benefits Atlas' landing page and structure. The Atlas includes: data reports on 11 co-benefits, 17 social economic factors, 382 local authorities; an interactive map; an example story; and methodology. }
  \label{fig:atlas-structure}
\end{figure*}

\subsection{Design workshop series}
\label{sec:workshops}
The 8 co-design workshops were structured into three stages---\textbf{Understand}, \textbf{Ideate \& Make}, and \textbf{Reflect}. Each stage addressed specific design questions and tailored activities. Each workshop lasted two hours, with a minimum one-week interval between consecutive workshops to allow time for reflection and preparation. The workshops were conducted in a hybrid format to accommodate remote participation. Workshop activities were facilitated through an online Miro whiteboard, while color pens and paper were provided for those attending in-person. 
In total, eleven participants took part in the design workshops (\cref{apx:workshop-attendance}): nine core team members who attended consistently across sessions (including one key stakeholder), and two additional external stakeholders who each joined a single workshop: one in W1 for their interest in atlas concepts, and one in W4 for their familiarity with local communities.
The first author led workshop documentation, including recordings, workshop materials, and participant attendance, and after each workshop produced a structured summary of emerging ideas and design directions.
These summaries were then presented and discussed in team reflection meetings, where they informed both prototype development and the preparation of subsequent workshops.
We outline the three-stage workshop setup as follows, with detailed documentation provided on our project website.

\textbf{Stage 1: Understand [W1\&2].} The first workshops aimed to create a shared understanding among core team members of the domain context and data, as well as potential audience groups and usage scenarios.
In the first workshop \jumpto{W1}, hosts shared familiar tools to establish domain context for the visualization experts and the external stakeholder, and discussed existing atlas examples taken from online atlas collections~\cite{wang_visualization_2025}. Collectively, we discussed our project goals and why we would need an atlas, i.e., whether an atlas was the appropriate tool for this project. 
In the second workshop \jumpto{W2}, we collectively created \textit{speculative} stakeholder persona cards, including potential professions, painpoints, and anticipated goals in using the Atlas. These mainly originated from discussions with the team of hosts, using their experience working on climate mitigation. From the persona cards, we speculated on stakeholders' potential data roles when using the Atlas, e.g., analyst, policy translator, defender of unpopular decisions.

\textbf{Stage 2: Ideate \& Make [W3-6].} The following four ideation workshops focused on different Atlas design dimensions, also described by Wang et al.~\cite{wang_visualization_2025}. These dimensions included 
\textit{atlas content page design}~\jumpto{W3}, 
\textit{visualizations}~\jumpto{W4}, 
\textit{transparency and methodology}~\jumpto{W5}, and 
\textit{entry points and visual consistency}~\jumpto{W6}. These workshops aimed to establish the visual, structural, and narrative framework of the atlas, i.e., what pages to include, what information these would feature, and how they were to be structured.
W3 asked \textit{what types of atlas pages are needed for different audiences?} based on the different page types identified in Wang et al.'s design space~\cite{wang_visualization_2025}, such as articles, dashboards, and reports. To that end, each participant picked one persona and scenario card from W2 and sketched atlas pages for these profiles, including visualizations, structure, and required explanations. 
In [W4], we worked towards specific analytical questions that visualizations could support, sketching and selecting appropriate data mappings and chart types. For example, we favored a bar chart over a violin plot for comparing co-benefit types within a local authority, prioritizing ease of interpretation for external stakeholders.
We did so through dedicated sketching sessions inspired by Kerzner et al's visualization opportunities workshops \cite{kerzner_framework_2019}. 
W5 involved sketching graphical labels~\cite{edelsbrunner2026visualization} alongside visualizations and page features to signpost contextual information such as different data types, methodological assumptions, data reliability, and key insights. This exercise clarified what the Atlas would need to disclose, and in what form: directly alongside visualizations, within summary texts, or on a dedicated methodology page.
In W6, we discussed how people could be onboarded and guided when using the Atlas. This workshop explored UI components such as navigation bars, search bars, and data overviews as prompts to brainstorm what entry points should be prioritized for the landing page's limited screen-space. Eventually, W6 focused on the visual identity and consistency of the Atlas. 
Here, Atlas hosts assigned color schemes and icons (e.g., for co-benefits and socio-economic factors) from an extensive color palette provided by visualization designers.

\textbf{Stage 3: Reflect [W7\&8].}  
During month 7 of the Atlas development, we conducted two additional workshops to reflect on implemented prototypes and identify emerging challenges.
In \jumpto{W7}, we reviewed early versions of page prototypes that contained real data. Domain experts were prompted to annotate page designs, verify patterns in visualizations and check that they corresponded to the main messages they had in mind, and to refine explanatory texts. 
The last workshop \jumpto{W8} focused on resolving outstanding design decisions. While page layouts and codebases had largely stabilized, several analytical and representational choices remained open, such as the default data measurements for co-benefits (per capita vs. total), and prioritization for certain socio-economic factors on the co-benefits page based on correlation strength. Visualization experts discussed these decisions with atlas hosts in order to draw on the hosts' domain knowledge and inform solutions.

\begin{figure*}[t!]
  \centering
  \includegraphics[width=1\textwidth]{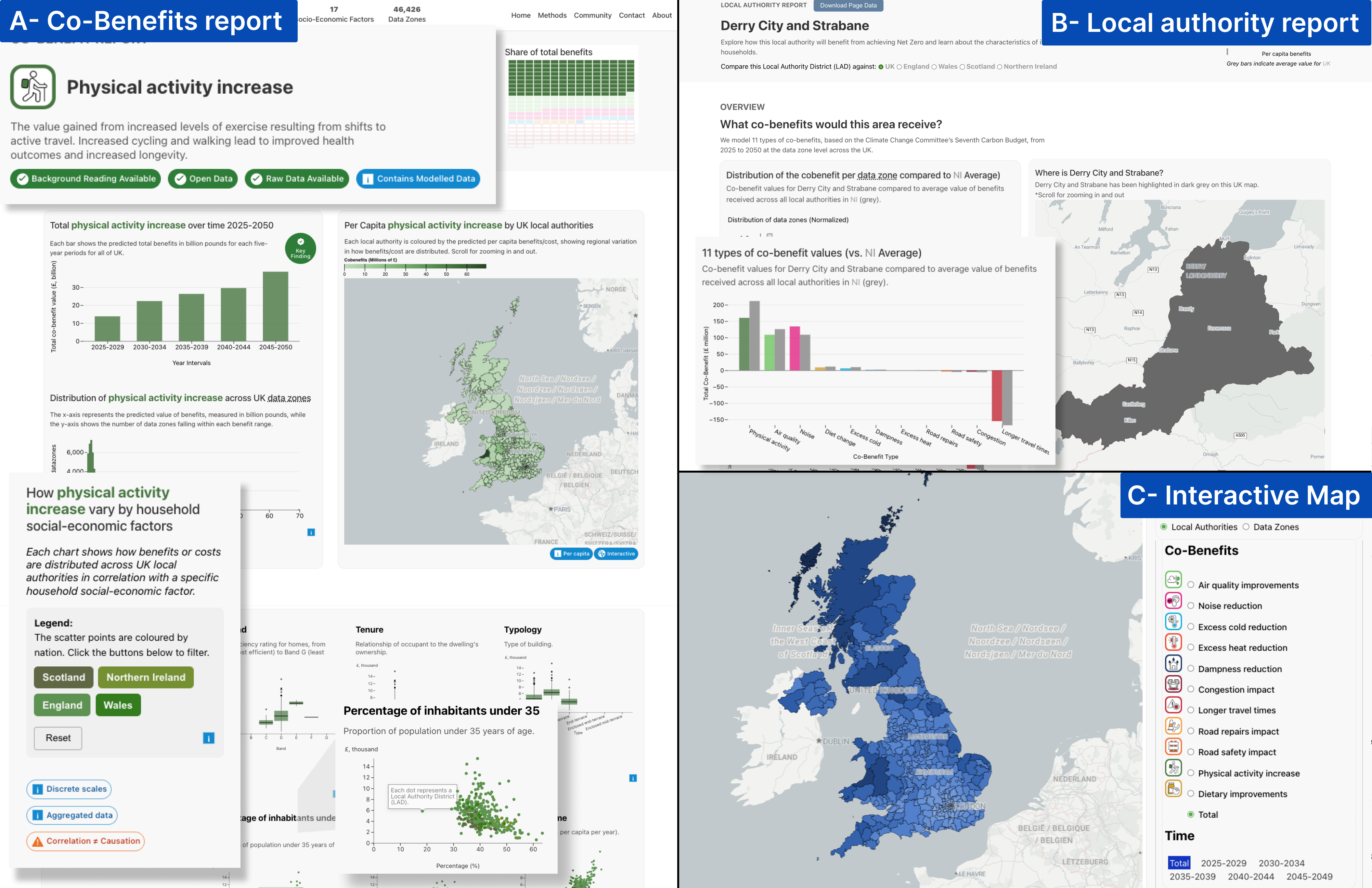}
  \caption{Examples of Atlas pages, including a co-benefit report page of physical activity increase (A), a local authority report for Derry City and Strabane (B), and the interactive map showing the geographical distribution of dampness reduction benefits (C).}
  \label{fig:atlas-pages}
\end{figure*}
\begin{table}[]
{\color{black}
\caption{Exemplary audience groups, goals, and tasks collected from design workshops.}
\vspace{-10pt}
\label{tab:tasks}
\centering
\begin{tabular}{p{\columnwidth}}
\midrule
\textbf{Example audience groups}                                                                                                              \\
MP, Policy maker in national government, Business consultant, Local authority Net-Zero officers, Local business owners...            \\ \midrule
\textbf{Example goals}                                                                                                                        \\
Promote climate action investment, defend unpopular local decisions, evaluate climate intervention outcomes, seek local funding, ... \\ \midrule

\textbf{Example tasks}                                                                                                                        \\
Compare co-benefits outputs in a local authority across data zones                                                                    \\
Compare co-benefits of the 6th carbon budget for local area versus UK                                                                 \\
Identify the roles of different co-benefits UK wide over time                                                                         \\
Compare the co-benefits impact on different types of households                                                                       \\
Compare SE factors across three local authorities                                                                                     \\
Understand different co-benefits for lower income areas      
                                        \\
Understand how co-benefits change over time for certain areas, ...       \\ \midrule
\end{tabular}
}
\vspace{-20pt}
\end{table}

\section{Atlas Design Decisions}
\label{sec:atlas}

The workshops yielded a range of specifications about 
(i) potential audience groups and their professional profiles, 
(ii) their high-level goals and
(iii) potential tasks for the atlas to support (\cref{tab:tasks}). Although not exhaustive, this list offers a useful starting point for identifying the diverse needs the atlas might support.
In this section, we present the features and design rationales of the UK Co-Benefits Atlas, informed by these findings. In addition to the workshops, these decisions evolved through a series of intermediary atlas iterations and reflection meetings (\cref{fig:project-overview}). We describe not only features that were ultimately included in the atlas, but also their alternatives and unrealized directions.

Prior work identifies the \textit{page} as one of several key design dimensions of visualization atlases~\cite{wang_visualization_2025}. In this context, a page (in the form of a web page) presents information about a specific data element or feature of the atlas. We introduce the UK Co-Benefits Atlas features by its landing page and a set of content pages that include 11 Co-benefits report pages (\cref{fig:atlas-pages}-A), each providing all the data available for the respective co-benefit, 382 local authority report pages showing data about city-sized communities (e.g., City of Edinburgh) (\cref{fig:atlas-pages}-B); 17 socio-economic factor report pages (e.g., unemployment rate, number of cars per household). The Atlas also features a landing page, an interactive map with all 46,426 Data Zones, a data story page, and a methodology page. Each report page features between 5-20 individual visualizations with titles and brief explanations. Pages are linked where necessary, e.g., to socio-economic factor and co-benefits pages through clickable text and visual marks, and to other local authorities and Data Zones via an interactive map~(\cref{fig:atlas-pages}-C). 

We also discussed additional page types that would support more specific tasks, such as temporal pages focusing on individual years and comparison pages between Data Zones and co-benefits \jumpto{W2}. We did not pursue these ideas due to the additional complexity they would introduce for navigation and the landing page (see below). However, these concepts highlight the potential of an atlas as a collection of pages dedicated to specific tasks (e.g., overview, analytics, explanation). Below, we detail the page types ultimately included in the atlas.

\paragraph{\textbf{Landing pages}} are the pages that users first encounter when visiting a visualization atlas. They aim to present the overarching ideas of the atlas while providing possible entry points into its content. They must provide context and explanations, while also communicating what users can learn, what the data are, and how to navigate into the details of the atlas. The challenge of designing landing pages is to include all this information within one page (limited screen space) in an inviting way. We decided to feature a `cover visualization' in the form of a waffle chart on the UK Co-Benefits Atlas landing page, showing the total amounts of individual co-benefits (see Fig.~\ref{fig:teaser}). The visualization aims to communicate \textit{a)} the data-driven approach of this platform, and \textit{b)} the main data findings \jumpto{W1}. As we use the same waffle chart design across several pages, it also acts as a distinctive visual landmark of the atlas.
Explanatory text introduces co-benefits and informs potential stakeholders about what the atlas is and how to use it. We also featured a prominent search bar (to jump to specific local authorities), links to data stories, and all report page types with preview thumbnails. 
These entry points were shaped together during workshops~\jumpto{W6} and were developed for two professional audiences \jumpto{W1}+\jumpto{W2}: those interested in specific climate interventions (e.g., housing retrofits) and those interested in specific geographic areas (e.g., city councils). 

\paragraph{\textbf{Content pages}} provide data, visualization, and information about concepts in the atlas or elements in the data. A major challenge in content page design was how to slice and group information into individual visualizations within and across different report pages. In early discussions about the dataset \jumpto{W1} and during the sketching activity \jumpto{W4}, we found different ways of presenting co-benefits data: different granularity of geographical distribution, total vs. per capita benefits, temporal distribution in relation to socio-economic factors, etc. 
Early workshops \jumpto{W2} also led to a range of speculative designs, including geographical comparison analysis, policy impact stories, customizable dashboards, a geographical data report, policy simulation, etc. However, without direct evidence of which stories different stakeholders would find most valuable, or how they would engage with the data in practice, it was difficult to commit to specific narratives. We therefore decided that each report should emphasize one primary data element~\jumpto{D2}, structured consistently from overview to detail. To scale this across the Atlas, we designed three bespoke \textbf{page templates} corresponding to the main data dimensions: co-benefits, local authorities, and socio-economic factors (see Fig.~\ref{fig:teaser}). These templates emerged naturally from the available data and workshop sketches~\jumpto{W4}. For example, a co-benefit page focuses on a single co-benefit and shows its national distribution (Fig.~\ref{fig:atlas-pages}.A), while a local authority page provides a zoomed-in view of a single location and covers all 11 co-benefit types (Fig.~\ref{fig:atlas-pages}.B).
Workshop insights~\jumpto{W4} further indicated a stakeholder's preference for individual charts, leading to the use of standard, self-contained visualizations (e.g., bar charts, maps) that can be easily interpreted and exported, rather than using more complex encodings.

While the report pages present information in a structured and communicative form, some stakeholders might need more analytical angles (e.g., community stakeholders focusing on data zone level interventions~\jumpto{W2}). An \textbf{interactive map} of the UK shows co-benefits for all 46k local authority Data Zones to maximize open-ended exploration, particularly for more advanced users ( Fig.~\ref{fig:atlas-pages}.C). Grounded in the geospatial nature of the dataset~\jumpto{W1}, the map offers a spatial lens on the data. An interactive legend allows users to define the co-benefits, socio-economic factors, and time ranges shown on the map. 

Complementing data exploration, the atlas provides \textbf{written articles}, currently comprising a policy brief with main insights and a methodology page (\cref{fig:atlas-structure}). 
The need for explanatory content was raised repeatedly~\jumpto{W1}\jumpto{W2}\jumpto{W5}, driven by the complexity of the data and the novelty of quantified co-benefits, underlying domain concepts, and the data modelling process.
We initially had planned workshops on creating more data stories~\jumpto{M6}, focusing on storytelling, chart annotations, infographics, and fully narrative articles~\cite{riche2018data}. However, we decided to table these stories as extendable future avenues. At the time of design, the data were simply too new and lacked in-depth analysis. 
Instead an example data story was co-crafted with a close stakeholder during later stages of the Atlas development and centered around a streamgraph from a local authority report (~\cref{fig:atlas-structure}-story), using it along with insights from the Atlas to illustrate the long-term returns of climate investment.
At the time of writing, however, updated stories with published reports using the Atlas data are available, such as the Scottish Government's Climate Change Plan 2026-2040 \cite{scottishgov2025climateplan}.

\section{Stakeholder discussions}
\label{sec:stakeholders}

The emerging UK Co-Benefits Atlas prototype described above allowed a first engagement with stakeholders external to the design process. We presented the Atlas to 25 such external stakeholders as part of dedicated onboarding sessions. We began these sessions as soon as the Atlas included a landing page and report pages for local authorities and co-benefits implemented with real data (Month~7). At this stage, any study would not qualify as an evaluation of the Atlas but as a means to assess some of the designs described in Section 4 and whether an atlas was a promising direction for trying to make the data in question accessible. Consequently, we were mostly interested in the following questions: 
\textit{a) What are people's \textbf{first impression} when engaging with the atlas and do they find it too complex?},
\textit{b) How do people \textbf{engage with the atlas} and what information do they look for?},
\textit{c) Which design features and characteristics cause \textbf{friction}?} and 
\textit{d) How do people \textbf{imagine using an atlas} for their work?} 

\subsection{Participants and session setup}

We organized four public outreach events to promote the Atlas and meet potential stakeholders. The events were advertised through organizational mailing lists and networks and attracted over 200 participants, including $\sim$20 colleagues within the core team’s institution, $\sim$50 local climate policy professionals, $\sim$100 officers from local governments across the UK, and  $\sim$50 members of community-led climate networks with diverse practitioners. Following the second outreach event, the core team started onboarding sessions to help potential users get familiar with the atlas and let them ask questions, discuss use cases, etc.
The onboarding sessions were promoted on the Atlas contact page, on LinkedIn, and announced during the outreach events. 
25 participants attended one of our onboarding sessions: 11 university staff providing professional services to local business development and government policy (P1-11); 2 staff from social enterprises (P11-12); 5 from local climate community hubs (P14-18), 2 from public libraries (P19-20), 2 from public consultancies (P21-22), and 3 holding different positions in government organizations (P23-25). 
\Cref{apx:external-stakeholders} details the external stakeholders' organizational affiliations and areas of expertise.
None of the participants indicated proficiency in data analysis or interpreting visualizations, despite their substantial professional expertise. 

\subsection{Data Collection \& Analysis}
We ran a total of 15 onboarding sessions (11 individual and four group sessions), each attended by 2-5 participants. 
This study is approved by Informatics School ethics committee's procedure (\#591184) in the University of Edinburgh.
The sessions were conducted online via Microsoft Teams and lasted around one hour each. The first author facilitated the sessions along with one of the Atlas hosts. 
During the sessions, participants were first asked to introduce their professional background and familiarity with co-benefits. This was followed by a walkthrough of the Atlas structure and page features. Participants were then invited to freely explore the Atlas and ask questions. 
We asked participants to think aloud to record their thoughts. Facilitators replied to questions, clarified information, and asked follow-up questions.

All onboarding sessions were recorded and automatically transcribed. The lead facilitator conducted a two-stage thematic analysis~\cite{braun2006using}. First, the transcripts were reviewed to identify information relevant to: participants' professional profiles, how they explored the Atlas report pages, and the usage scenarios they proposed. During this stage, ad-hoc thematic tags were created and applied to conversation segments to capture both the topic discussed (e.g., data credibility) and where it arose in the Atlas (e.g., local authority report). In the second round, these tags were reviewed, refined, and merged into broader categories. The thematic analysis was mainly conducted by the first author; emerging themes were discussed within the core team~\jumpto{M22}, who collectively reflected on and helped refine the categories to identify key barriers and usage scenarios. The final themes reported below reflect this shared review.

\subsection{Observations}

All participants expressed enthusiasm for the Atlas and its features. We also observed engagement with individual Atlas pages, i.e., participants reading and trying to interpret individual visualizations.

\subsubsection*{\textbf{First Impressions---What are people's first impressions of engaging with the Atlas?}}
For many participants, the Atlas \textbf{looked impressive yet overwhelming} at first glance. While participants quickly grasped the richness of content in the Atlas, the represented data was experienced as complex, even overwhelming: \itquote{it looks like there's a lot of helpful information---would be my initial reaction. But then, I wouldn't know where to start} (P16). 
Participants also asked for similar onboarding sessions in the future to  \itquote{gain my confidence and navigate the Atlas} (P20).

\vspace{-5pt}
\subsubsection*{\textbf{Engagement---How do people explore the atlas? What information do they look for?}}

Participants had \textbf{different approaches to exploring the Atlas}, often guided by their personal experience or professional background. We observed explorations of the low-level Data Zones that led to ideas around community-level applications: \itquote{If an [interest] group wants to apply for [a specific intervention], they can go straight into that hyper-local area and see if we get money to install this} (P15). 
Other participants started with specific economic factors, such as housing energy bands (P12) or the inhabitants' age profile (P20). 
This supports the idea that atlases accommodate a range of tasks and exploration approaches and suggests the importance of providing various exploratory features and diverse points of view on the data. Participants did not necessarily follow the `overview-then-detail' journey~\cite{luciani2018details}, and often began their exploration from a feature they were interested in or involved with.

Participants also \textbf{asked specific questions related to the data and methodologies}, for example, concerning the measures used in the modeling process: \itquote{does it know in a Data Zone which properties are suitable for certain types of retrofit measures?}~(P12). They also highlighted the importance of methodology: \itquote{you don't have to get the great details, but just give me a sense of what that is, the parameters, the assumptions and the methodology}~(P14).
The Atlas triggered more in-depth methodology questions as sessions progressed, especially by those with an academic background. For example, P5 asked about the interconnection between societal impact data \itquote{how do you set the boundary of the co-benefits?}. We also frequently observed instances of participants interpreting but also questioning the presented information based on their work experience. For example, P15 mentioned certain interventions \itquote{are not going to land greatly in the city where I work.} 

Strongly related, we found participants \textbf{curious about data not included in the Atlas}. For example, participants asked for clarifications on why certain co-benefits and socio-economic factors were not included in the Atlas: \itquote{so are you only using data sets that apply to the whole of the UK or do you use regional data?} (P9). This suggests a keen interest and critical reflection on methodological limitations and trying to interpret the Atlas in its wider context, i.e., atlases as platforms that \textit{could} include more data. Some participants were particularly skeptical---in a positive sense---about the data, e.g., when local data contradicted their day-to-day experience: \itquote{[For local travel times,] you might want to have a little asterisk and say this is not the case necessarily for everyone} (P6). 
Some participants suggested possible approaches to address their concerns, for example via the introduction of external datasets. For example: \itquote{Did you use past history data and compare it to see like how accurate the future data would be?}~(P22).

\subsubsection*{\textbf{Frictions---Which content and design features present challenges in exploration and interpretation?}}

The methodological questions outlined above represent critical reflections that can be considered as frictions between participants' understanding (and belief) in the data presented. Moreover, we also found that some of the data representations presented challenges for interpretation. For example, participants found it challenging to \textbf{understand the quantification of abstract co-benefits}: i.e., how social benefits can be compared, \itquote{I'm not sure how you compare that value of my [travel] time relating to the value of breathing cleaner air} (P6).  While this is related to methodology, P9 acknowledged the complexity of \itquote{the [multi-faceted] data that you have to put in[to the Atlas].} We believe that this challenge of explaining analytical values and measures to describe complex real-world topics is inherent in the comprehensive approach of visualization atlases in general ( \cite{wang_visualization_2025}). 

The interpretation of patterns in the visualizations and understanding their real-world meaning also presented challenges. While participants found the standard visualizations like bar charts easy to read, some found it challenging to \textbf{translate observed patterns to practical policy implications}. For example, P6 asked about the disproportionate contributions across the 11 co-benefits categories (\autoref{fig:atlas-pages}-B),\itquote{Why will physical activity go up that much?}
This lack of clarity may negatively impact participants' willingness to take action: \itquote{So if I talk to a policy maker, improved diets sounds great, but how? What specifically do you need me to do?}~(P5).
Related to this are frictions imposed by participants questioning the \textbf{reliability of the data}. While participants understood that projection data does not promise certainty, they still sought some indication of the data's reliability to represent reality: \itquote{What can one do to avoid overstating what is possible? Obviously you can't have a detailed survey of every single property!}~(P12). While the complete data reliability explanations exist in our methodology page, it is a challenge to surface and deliver targeted explanations that suits different stakeholders' requirements.

5 participants found it challenging to \textbf{maintain data context when switching pages} in the Atlas, navigating between pages and visualizations. For example, when switching to a local authority page, P17 asked about the time range shown in the co-benefits distribution: \itquote{So what is the year that it starts in your graph? What is the year model 2025?} Similarly, P19 noted \itquote{the landing page figure (waffle), is that national as well or can you change that by district or region?}. We believe this to be a general challenge in visualization atlas design, especially if many pages contain similar (looking) visualizations and information. 

Last but not least, participants asked for \textbf{proofs of integrity} from some form of authority, e.g., by displaying logos from the core team's institutions alongside a clear statement that the methodology aligns with (UK) government guidance on policy evaluation. The lack of such links challenged the trust that participants have in the information provided and actions they may take. For example, one participant noted that they needed to know  \itquote{where the funding comes from and what the future is for the Atlas. This will help me decide how comfortable I feel putting resources into it}~(P18). Some participants further expressed a strong interest in quoting a credible statement about whether the Atlas had been recognized or officially adopted by local governments. Establishing trust is key for visualization atlases to have an impact.

\begin{figure*}[ht!]
  \centering
  \includegraphics[width=1\linewidth]{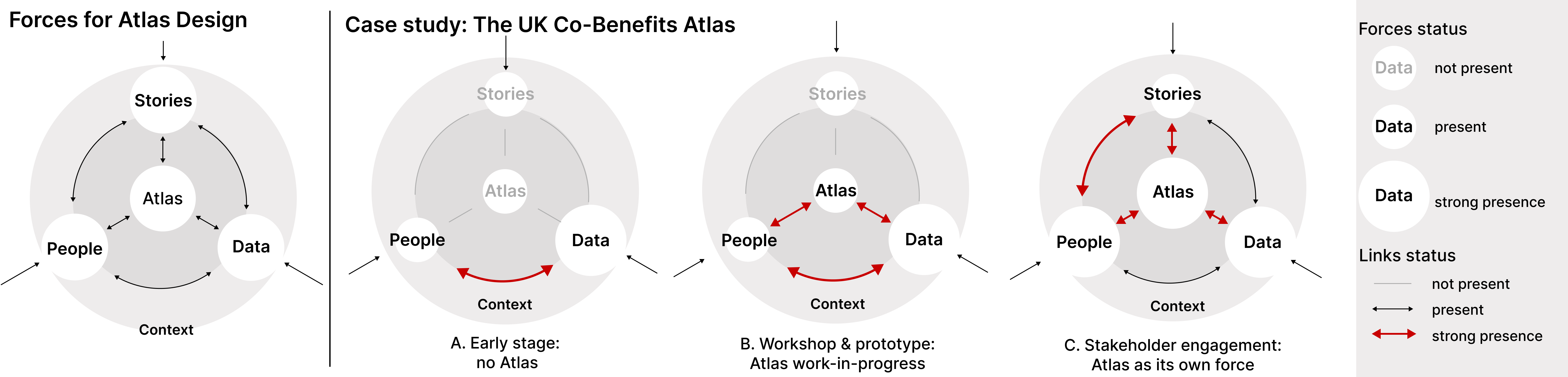}
  \vspace{-10pt}
  \caption{Navigating shifting forces driving atlas design: data, stories, people, atlas, and context. We illustrate how these forces changed in our case study from A) early stage, B) workshop \& prototype, to C) stakeholder engagement.}
  \label{fig:forces}
\end{figure*}

\subsubsection*{\textbf{Usage scenarios---How do people envision using an atlas in the long run?}}
\label{sec:stakeholder-scenarios}

For our participants, the onboarding sessions were their first in-depth exploration in the UK Co-Benefits Atlas, however, these guided explorations remain limited to show long-term atlas usage. Nevertheless, participants highlighted the Atlas as a novel and potentially valuable source of information and expressed enthusiasm for sharing the Atlas with peers in their professional and community networks, \itquote{I'd love to show our librarians and they can demonstrate this to their library communities} (P19). Participants working with local communities acknowledged the value of publicly accessible research resources:\itquote{We wouldn't be able to do this---we don't have the capacity or the required skills with it being at a national level and linked to the institution, which gives it credibility that we could use in funding applications or in conversations with the council}~(P16). Participants mentioned four potential usage scenarios, ranging from lightweight engagements, such as learning, to stronger impacts through decision-making.

\textbf{Scenario 1: Learning about co-benefits and co-benefits data}---As co-benefits still remain an emerging concept, learning about the scope of the Atlas and its core data dimensions represents an early-stage use, where the Atlas would function not primarily for accessing data but for learning. Six participants signed up for training sessions out of curiosity rather than with a concrete application in mind, for example: \itquote{I was keen to understand in a bit more detail exactly how it’s built and how it works, for future reference} (P7).

\textbf{Scenario 2: Raise awareness and foster wider engagement}---The Atlas was found to have the potential to reach wide audiences in different ways. 
P19, a librarian, emphasized promoting information in a non-intrusive way, stating that \itquote{we try not to force things on people, we like people to find their own feet, especially when it can be classed as a sensitive issue, such as climate change.}
Several participants viewed the Atlas as a valuable resource for policy makers, for example P6 shared \itquote{it's important that we can bring resources like this to national government, make them aware of things that are available.} 
Finally, participants discussed how to introduce the Atlas, including how to communicate \itquote{the importance and the uncertainties around those assumptions} (P9), and convey the take-home message: \itquote{what we're saying [now] is the investment and the costs, but these co-benefits in 5 or 10 years time will be much more tangible} (P9).

\textbf{Scenario 3: Justify \& plan decision making}---Participants envisioned the Atlas to be useful in different ways to support decision making.
Seven participants wanted to use the Atlas for seeking confirmation and evidence for predefined agendas, such as justifying the positive social impact of prior work. For example: \itquote{So we want to try and convince the government to fund this long term to 2045 [...] this Atlas could be invaluable to convince them}~(P14). However, the Atlas also challenged participants' ideas. For example, P1 was shocked at the negative outcome of road conditions and expressed concerns about the related transport policy in place, raising the critical question of how public data influence decision-making.
Finally, participants also underscored the Atlas' power to guide future actions and priorities. As P14 pointed to promising co-benefits in physical activity increase that are closely linked to active transport policies, saying the Atlas changed their perspective. 
P25 further highlighted the Atlas can support the allocation of limited resources: \itquote{We want to pick actions that [...] maximize return of co-benefits on investment}. 
\section{Reflecting on Atlas Design}
\label{sec:discussion}

\newcommand{\data}{\textsc{Data}\xspace}
\newcommand{\people}{\textsc{People}\xspace}
\newcommand{\stories}{\textsc{Stories}\xspace}
\newcommand{\atlas}{\textsc{Atlas}\xspace}
\newcommand{\context}{\textsc{Context}\xspace}
\newcommand{\force}[2]{#1$\rightarrow$#2}
\textit{What can we learn from designing the UK Co-Benefits Atlas for future atlas design?} To answer this question, we examined our design process in retrospect along the following questions: \textit{What information was available to the design team and when? Which conditions or people influenced design decisions? What constrained the progress? How would we approach atlas design with the knowledge and experience we have today?} Our reflections led us to consider atlas design applying the metaphor of \textit{``forces''}, describing major influences that shape the design process at different moments. The forces we observed in our process are \data, \people, \stories, \atlas, and \context (\cref{fig:forces}). 
In this section, we first introduce the forces, then draw on other atlas cases to illustrate how this lens can be used to understand different atlas processes, and finally discuss implications for future atlas design.

\subsection{Forces in the atlas design process}
The \textbf{\data force} is made up of the available data, their size, complexity and comprehensiveness, uncertainties, and analysis processes. It also captures the data origin, their construction and technologies involved. 
In our case, the co-benefits data were provided prior to the design process by the Atlas hosts, heavily influenced by government guidelines.
\data was a relatively dominant force from the beginning for our case, where \data dimensions influenced sketching activities (\cref{sec:workshops}, \jumpto{W4}), and \data complexities challenged the resources required to construct visualizations and explanations.

The \textbf{\people force} encompasses all actors involved in or engaging with the visualization atlas. This includes atlas hosts who commission the atlas, provided the data and visualization designers. It also includes external stakeholders that the atlas is aimed at, such as the participants in our design workshops and onboarding sessions. 
\people are a force shaped by their knowledge of the topic; their interests and tasks for the data; and their time and willingness to engage with, trust, promote, and contribute to the atlas (\cref{fig:forces}-A).
During the workshop and development stages, the \people force was shaped by the vision of the atlas hosts, input from selected external stakeholders, and the designers’ ability to translate this vision into concrete representations. 
Given the range and uncertainty around potential audiences and tasks, we deliberately used these workshops to brainstorm and amplify wider stakeholder perspectives as design opportunities.

Through the continued interaction between \people and \data, the \textbf{\atlas} gradually emerged as its own force (\cref{fig:forces}-B), taking shape through its structure, visualizations, navigation, and explanatory elements. 
This Atlas is now a dense yet highly structured platform with a landing page, over 400 content pages for different data elements (co-benefits, local areas, socio-economic factors), an interactive map, and stories. 
Once established, this Atlas became an active force feeding into interactions between \people and \data by shaping how data can be accessed and interpreted. In particular, it enabled engagement with external stakeholders by providing a concrete medium through which to explore and interpret the data (\cref{sec:stakeholders}). 
While the Atlas may be perceived as overwhelming at first, the breadth of data was regarded as a valuable resource across use cases such as learning, awareness, decision-making, and future planning. Stakeholders sought specific information, critically interrogated data methodologies, and expressed their interpretation of complex co-benefits concepts.

This led us to define \textbf{\stories}, which mediates interactions between people and data through the atlas (\cref{fig:forces}-C). \stories can include anything from chart annotations and explanations to methodology documents, policy use cases\footnote{\url{https://ukcobenefitsatlas.net/stories/story.pdf}}, and, eventually, high-level narratives about climate mitigation. 
Stories mainly emerged from \people's analysis of the data, and can highlight the relevance of the atlas, its topic and its wider context. 
In our case, stories are intended to originate more strongly from external stakeholders than from the hosts, positioning \stories as a force that becomes more prominent in later stages of development.
Since reaching a prototype and involving more stakeholders, developing more stories has the potential to constitute important entry points to the Atlas, helping connect with wider audiences (\cref{sec:stakeholders}).

We recognize \textbf{\context} as a final force which includes the environment in which the atlas design process takes place. It includes the atlas' topic (e.g., co-benefits for climate mitigation) and the public discourse around the topic which mirrors and influences its relevance to \people and \data as well as frameworks and sources for data collection or creation (e.g., government guidelines, explicit rigorous practices, quality control). The \context also includes pragmatic aspects such as funding, timelines, and external frameworks and guidelines for design. As a `background' force, shifts in \context can have a direct impact on data, stakeholders, stories and the atlas. We experienced such a shift which triggered \data change and consequently the \atlas design. During the development stage, the UK government decided to remove a particular data dimension (comparison between different futures) from its agenda. We consequently had to remove an entire data dimension from the Atlas, including some potential stakeholders' goals (\people).

\subsection{Understanding atlas design through forces}
The core idea of interacting forces can be a productive lens that informs visualization atlas design processes by spatially outlining possible design steps and challenges at a given point in time, rather than suggesting an iterative yet linear way of navigating atlas design.  
Revisiting interviews with atlas designers from our prior work~\cite{wang_visualization_2025}, we examined the forces present in other atlas projects. 

We first examined atlas cases that, like ours, have complex \data early on. For example, the Economic Complexity Atlas~\cite{GrowthLab2013Atlas} was grounded in data developed by economists, but also shaped by static visualizations previously published in books. These visual forms acted as an initial \atlas prototype. That project then transformed these prototypes into more comprehensive exploratory tools. Later, as policy audiences faced barriers in interpreting the visualizations and methodology (\people force becoming more prominent), the atlas started integrating more motivating explanatory pages and feature-walkthrough tutorials. The same team later published  \textit{Metroverse}~\cite{Metroverse}, an atlas on a related topic in economic analysis. Yet because its \data and \context differed substantially from the earlier atlas, Metroverse emerged as a distinct atlas rather than a direct continuation. This case illustrates how a new atlas may evolve as a result of forces shifting and transforming each other.
Still other atlases may follow more different processes due to different starting points. For example, the Atlas of \textit{Sustainable Development Goals (SDG)}~\cite{atlasSustainability} can be seen as \textit{story-driven}~\cite{wang_visualization_2025}, with internal experts leading the narrative of polished data-driven \stories~\cite{segel2010narrative}.
Here, design focuses less on fully representing comprehensive \data or supporting \people's analytical tasks, but more on optimizing visualizations, interactions, and page styles that communicate \stories effectively. As part of the SDG Atlas series, it is also shaped by \atlas issued previously, such as visual conventions and organizational patterns. Conversely, the \textit{Rare Diseases Observatory}~\cite{RareDiseasesSlovenia} began without consolidated \data and took a user-driven approach, with target audiences (\people) guiding data-source and visualization selection. The atlas began with initial use cases and data sources, but was designed to grow as more sources became available.

These examples, alongside our case study, illustrate the different ways in which atlases can take shape. The flexibility of forces to describe different project constellations makes them a useful lens to navigate atlas design and to enable reflection during and after visualization atlas design processes. We do not claim a gold standard methodology for matching particular force constellations with specific design practices. Instead, we argue that forces as a `lens' can serve as a useful conceptual tool for future atlas designers that can inform methodological choices in the design process as well as atlas solutions.

\subsection{Implications for atlas design}
These observations and reflections from our case as well as accounts about other atlases helped us identify a first set of implications for atlas design (G1-G4). 
Many of these guidelines may not appear novel on a general level, but are grounded in our hands-on experience and 
address atlas-specific challenges.

\textbf{G1: Embrace the fuzziness and breadth of audience goals and tasks.}
In our case study, the novelty of the co-benefits \data meant that target audiences, goals, use cases remained uncertain until a working prototype enabled us to showcase the data and engage potential audiences. Such uncertainties are common in visualization atlases intended for diverse audiences~\cite{IPCC, DatawheelUSA}. 
Established design study methodologies~\cite{sedlmair2012design, oppermann2020data, munznerNestedModelVisualization2009} centered on task abstraction remain useful, but need adaptation to accommodate a breadth of possible tasks rather than reducing them to a fixed task set.
For atlases, the ``fuzzy front-ends''~\cite{sanders2008co} include not only visualization, interaction and data~\cite{wei2026fuzzy}, but also uncertainties in audiences and tasks.
However, there can be a huge potential in open and fuzzy audiences. For example, in our process, embracing fuzziness meant including selected stakeholders (\people) in early ideation and speculating with them about potential audiences and use cases, for example, through persona building~\cite{pruitt2003personas}. Driving the design process forward despite missing information meant that we could create an atlas prototype that would later enable us to explore these uncertainties. It also allows making quick or entirely novel design decisions and seeking the audience these decisions appeal to.

\textbf{G2: Break down atlas design dimensions.} 
Visualization atlases involve multiple design dimensions that include page types, entry points, navigation, onboarding, etc~\cite{wang_visualization_2025}. In our case, these challenges in page composition and organization were more central to developing novel visualization techniques for complex tasks~\cite{brehmer2013multi, munznerNestedModelVisualization2009}. 
We used a series of workshops, adapted from visualization workshops~\cite{kerzner_framework_2019} and physicalization~\cite{huron_lets_2017}, to break the complex atlas design space into smaller, meaningful atlas-specific activities that participants (\people) could engage with more confidently and meaningfully.
While design dimensions can be unstable given the scale and volume of an atlas, the forces lens can help atlas teams reflect on available resources, understand prominent challenges at any stage, and select appropriate methods.

\textbf{G3: Maintain \people's continuous involvement with flexible approaches.} 
Prior work notes that sustained involvement is a common challenge as many projects lack the time or space to engage broader groups throughout~\cite{wang_visualization_2025}.
Echoing problem--solution co-evolution in designers' practice~\cite{sanders2008co, parsons2025beyond}, \people's involvement shaped how our atlas design evolved: early on stakeholders helped co-create the problem space by brainstorming possible goals and tasks, which became starting points for ideating \atlas design opportunities.
Once a prototype enabled broader engagement, onboarding sessions and public presentations gathered wider feedback to iterate features, develop data stories, and respond to stakeholder needs. Therefore, we propose that atlas design can switch and adapt participatory methods based on design tasks and available contributors.

\textbf{G4: Design for change.}
Visualization atlas projects are too complex for all outcomes to be foreseen from the beginning. As forces change, atlases need to respond flexibly across multiple design dimensions such as pages, navigation, and onboarding. This softens the stability and authority implied by the term ``atlas'', positioning atlas artefacts more as provisional ``immutable mobiles''~\cite{Latour_1986} or ``sandcastles''~\cite{hinrichs_defense_2019} that act as \textit{``aesthetic provocations''} and \textit{``mediators of ideas''}. In our case, the current structure supports certain changes, such as updating visualizations with new \data or adding report components and stories. More dramatic changes in \context, however, such as data methodology or atlas purpose, may require rethinking the structure itself such as in the case of Metroverse~\cite{Metroverse}. A sustainable atlas should thus absorb anticipated changes while clearly scoping when redesign becomes necessary.
\\

\section{Conclusion \& future work}
\label{sec:conclusion}

We have reported on a first-hand design process of a visualization atlas for a climate mitigation dataset. Through eight design workshops and iterative development we have developed a comprehensive visualization Atlas that includes over 400 data reports, an interactive map, and supporting articles on methodology and data stories. Reflecting on this process, also based on insights from stakeholder engagement through 15 onboarding sessions, we propose the metaphor of forces to understand and navigate the design process of visualization atlases. These forces---in our case defined by \data, \people, \stories, \context, and the \atlas itself act as a reflective and generative lens that frames atlas design as a spatial rather than linear process, where shifting forces influence the design process and each other as part of this. Based on our case study and other atlas examples, we discuss how different force constellations can emerge and link these constellations to different design methods. Our four design implications offer concrete considerations on navigating visualization atlas design without being prescriptive.
However, as visualization atlases are still an emerging area, more work is needed to understand their design. 
We call for more reporting of atlas design cases with different domains, starting points, and organizational contexts. Our forces can provide some theory to approach this reporting. They can help structure the design process or define constraints and priorities, i.e., limits to the extent a force is allowed to influence the design. They can help set priorities and find `the right moment' to make design decisions, e.g., involving new people, including (or removing) data, refining stories, etc.

Studying the long-term use and societal impact of atlases also remains key for future work. Since our Atlas launched in July 2025, Google Analytics reports over 4,200 active users as of March 2026. The Atlas has so far supported work by the UK Department for Energy Security and Net Zero, analysts involved in Scotland's Climate Plan~\cite{scottishgov2025climateplan}, Glasgow's climate action planning, and Manchester's work on financing climate action. These examples suggest that atlases can support diverse agendas, from policy-making and government decision support to business planning, advocacy, education, and public engagement. Future work therefore has to examine how broad audiences engage with atlases over time, including how they can reach groups often excluded from professional or policy-facing data discourse. Specifically, we need to understand whether and how atlases shape public discourse around complex real-world issues, and what responsibilities designers have when creating open platforms that may influence interpretation, debate, and decision-making.

\acknowledgments{%
	\textit{The UK Co-Benefits Atlas was funded by Scotland Beyond Net Zero seed funding. We thank all stakeholders who participated in workshops and onboarding sessions for their valuable contributions to this project. We also thank the reviewers for their constructive feedback, and our colleagues for their help with proofreading.
 }
}

\bibliographystyle{abbrv-doi-hyperref}

\bibliography{ref}

\newpage
\appendix
\crefname{appendix}{Appendix}{Appendices}
\Crefname{appendix}{Appendix}{Appendices}
\crefalias{section}{appendix}
\counterwithin{figure}{section}
\counterwithin{table}{section}
\renewcommand{\thefigure}{\thesection\arabic{figure}}
\renewcommand{\thetable}{\thesection\arabic{table}}
\clearpage
\newpage
\begin{center}
\large
APPENDIX OF\\[1mm]
\Large\noindent
\textbf{\textsf{
Designing a Visualization Atlas: Lessons \& Reflections from The UK Co-Benefits Atlas for Climate Mitigation
}}\\[2mm]

\normalsize
Jinrui Wang$^1$, Alexis Pister$^2$, Sian Pillips$^2$, Sarah Bissett$^1$,\\
Ruaidhri Higgins-Lavery$^1$, Clare Warmby$^1$, \\
Andrew Sudmant$^1$, Uta Hinrichs$^1$, Benjamin Bach$^{1,3}$\\[1mm]

$^1 $The University of Edinburgh, UK\\
$^2 $City St George's, University of London, UK\\
$^3 $Inria, France
\normalsize
\end{center}

\section{Design workshop attendance}
\label{apx:workshop-attendance}
\begin{xtabular}{lllll}
\hline
\textbf{Role} & \textbf{Core Team} & \textbf{Workshop Attendance} & \textbf{Organization} & \textbf{Expertise} \\
\hline
Vis designer & true & 8 {[W1, W2, W3, W4, W5, W6, W7, W8]} & University & Vis research, workshop planning, design \\
Vis designer & true & 7 {[W1, W2, W3, W4, W5, W7, W8]} & University & Engineering, design \\
Vis designer & true & 4 {[W5, W6, W7, W8]} & University & Engineering, design \\
Vis designer & true & 1 {[W8]} & University & Vis research, workshop planning \\
Vis designer & true & 6 {[W1, W2, W3, W4, W5, W7]} & University & Vis research, workshop planning, design \\
Atlas host & true & 7 {[W1, W2, W3, W4, W5, W7, W8]} & University & Socio-economic climate research \\
Atlas host & true & 8 {[W1, W2, W3, W4, W5, W6, W7, W8]} & University & Data modeling and analysis \\
Atlas host & true & 8 {[W1, W2, W3, W4, W5, W6, W7]} & University & Policy communication \\
External stakeholder & true & 6 {[W1, W2, W3, W4, W5, W7]} & University & Data-driven climate policy \\
External stakeholder & & 1 {[W1]} & University & Energy sustainable transition \\
External stakeholder & & 1 {[W4]} & University & Climate community communication \\
\hline
\end{xtabular}

\section{Onboarding session participants}
\label{apx:external-stakeholders}
\begin{xtabular}{llll}
\hline
\textbf{Role} & \textbf{Participant ID} & \textbf{Organization} & \textbf{Expertise} \\
\hline
External stakeholder & P1 & University & business engagement \\
External stakeholder & P2 & University & energy retrofit research \\
External stakeholder & P3 & University & community engagement \\
External stakeholder & P4 & University & sustainability management \\
External stakeholder & P5 & University & energy policy \\
External stakeholder & P6 & University & government climate policy \\
External stakeholder & P7 & University & government climate policy \\
External stakeholder & P8 & University & government climate policy \\
External stakeholder & P9 & University & government climate policy \\
External stakeholder & P10 & University & government climate policy \\
External stakeholder & P11 & University & government climate policy \\
External stakeholder & P12 & Social enterprise & housing retrofit \\
External stakeholder & P13 & Social enterprise & business engagement \\
External stakeholder & P14 & Local climate hubs & local actions \\
External stakeholder & P15 & Local climate hubs & local actions \\
External stakeholder & P16 & Local climate hubs & local actions \\
External stakeholder & P17 & Local climate hubs & local actions \\
External stakeholder & P18 & Local climate hubs & local actions \\
External stakeholder & P19 & Public library & information access \\
External stakeholder & P20 & Public library & information access \\
External stakeholder & P21 & Public consultancy & government policy planning \\
External stakeholder & P22 & Public consultancy & local charities \\
External stakeholder & P23 & Government & sustainability management \\
External stakeholder & P24 & Government & data analysis \\
External stakeholder & P25 & Government & local sustainability policy \\
\hline
\end{xtabular}

\end{document}